\documentclass[aps,
preprint,superscriptaddress,groupedaddress]{revtex4} 
\usepackage{latexsym}       
 \usepackage{natbib}
\usepackage{graphicx}  
\usepackage{anysize} 
\usepackage{array}
\usepackage[T1]{fontenc}           
\begin{document}

\title{Epidemic Model with Isolation in Multilayer Networks}

\author{L. G. Alvarez Zuzek} \email{lgalvere@mdp.edu.ar} \affiliation{Departamento de F\'{i}sica,
  Facultad de Ciencias Exactas y Naturales, Universidad Nacional de
  Mar del Plata, and Instituto de Investigaciones F\'{\i}sicas de Mar
  del Plata (IFIMAR-CONICET), De\'an Funes 3350, 7600 Mar del Plata,
  Argentina}
\author{H. E. Stanley} \affiliation{Center for Polymer
  Studies, Boston University, Boston, Massachusetts 02215, USA.}
\author{L. A. Braunstein} \affiliation{Departamento de F\'{i}sica,
  Facultad de Ciencias Exactas y Naturales, Universidad Nacional de
  Mar del Plata, and Instituto de Investigaciones F\'{\i}sicas de Mar
  del Plata (IFIMAR-CONICET), De\'an Funes 3350, 7600 Mar del Plata,
  Argentina}\affiliation{Center for Polymer Studies, Boston
  University, Boston, Massachusetts 02215, USA.}

\begin{abstract}

The Susceptible-Infected-Recovered (SIR) model has successfully
mimicked the propagation of such airborne diseases as influenza A
(H1N1). Although the SIR model has recently been studied in a
multilayer networks configuration, in almost all the research the
isolation of infected individuals is disregarded. Hence we focus our
study in an epidemic model in a two-layer network, and we use an
isolation parameter $w$ to measure the effect of quarantining
infected individuals from both layers during an isolation period
$t_w$. We call this process the
Susceptible-Infected-Isolated-Recovered (SI$_{\rm I}$R) model. Using
the framework of link percolation we find that isolation increases the
critical epidemic threshold of the disease because the time in which
infection can spread is reduced. In this scenario we find that this
threshold increases with $w$ and $t_w$. When the isolation period is
maximum there is a critical threshold for $w$ above which the disease
never becomes an epidemic. We simulate the process and find an
excellent agreement with the theoretical results.

\end{abstract}

\maketitle

\section*{INTRODUCTION}

Most real-world systems can be modeled as complex networks in which
nodes represent such entities as individuals, companies, or computers
and links represent the interactions between them. In recent decades
researchers have focused on the topology of these networks
\cite{Bara_rew}. Most recently this focus has been on the processes
that spread across networks, e.g., synchronization
\cite{Lar_09,Ana_01}, diffusion \cite{Pan_04}, percolation
\cite{Dun_01,Coh_03,New_03,Val_11}, or the propagation of epidemics
\cite{New_05,past_01,Buo_13,past_02,Gra13,Cozzo_13,Mar_11,Sanz_14,Sahneh_14}.
Epidemic spreading models have been particularly successfully in
explaining the propagation of diseases and thereby have allowed the
development of mitigation strategies for decreasing the impact of
diseases on healthy populations.

A commonly-used model for reproducing disease spreading dynamics in
networks is the susceptible-infected-recovered (SIR) model
\cite{Ander_91,Bai_75}. It has been used to model such diseases as
seasonal influenza, such as the SARS \cite{Col_07}. This model groups
the population of individuals to be studied into three compartments
according to their state: the susceptible (S), the infected (I), and
the recovered (R). When a susceptible node comes in contact with an
infected node it becomes infected with an intrinsic probability
$\beta$ and after a period of time $t_r$ it recovers and becomes
immune. When the parameters $\beta$ and $t_r$ are made constant, the
effective probability of infection is given by the transmissibility
$T= 1-(1-\beta)^{t_r}$ \cite{Dun_01,Coh_hand}.

As infected individuals cannot be reinfected, the SIR model has a
tree-like structure with branches of infection that develop and
expand. Because in its final state this process can be mapped into
link percolation \cite{Bra_07,New_03}, we use a generating function to
describe it. In this framework, the most important magnitude is the
probability $f$ that a branch of infection will expand throughout the
network, \cite{Bara_rew,Bra_07}. When a branch of infection reaches a
node with $k$ connections across one of its links, it can only expand
through its $k-1$ remaining connections. It can be shown that $f$
verifies the self-consistent equation $ f=1-G_1(1-Tf)$, where
$G_1(x)=\sum_{k=k_{\rm min}}^{k_{\rm max}}kP(k)/\langle k \rangle x^{k-1} $ is
the generating function of the underlying branching process
\cite{New_03}. Note that $G_1(x)$ here represents the probability that
the branches of infection will not expand throughout the network. At
the final state of this process, the branches of infection contribute
to a spanning cluster of recovered, previously infected
individuals. Thus the probability of selecting a random node that
belongs to the spanning cluster is given by $R=1-G_0(1-Tf)$, where
$G_0=\sum_{k=k_{\rm min}}^{k_{\rm max}}P(k)x^k$ is the generating function of
the degree distribution. When $T \leq T_c$ there is an epidemic-free
phase with only small outbreaks, which correspond to finite cluster in
link percolation theory. But, when $T > T_c$ an epidemic phase
develops. In isolated networks the epidemic threshold is given by $T_c
= 1/ (\kappa-1)$, where $\kappa$ is the branching factor that is a
measure of the heterogeneity of the network. The branching factor is
defined as $\kappa\equiv\langle k^2 \rangle / \langle k \rangle$,
where $\langle k^2 \rangle$ and $\langle k \rangle$ are the second and
first moment of the degree distribution, respectively.

Because real-world networks are not isolated, in recent years
scientific researchers have focused their attention on multilayer
networks, i.e., on ``networks of networks''
\cite{Bul_01,jia_02,Gao_12,Gao_01,Val13,Bax_01,Bru_01,Brummitt_12,Lee_12,Gomez_13,Kim_13,Cozzo_12,Car_02,Kal_13}. In
multilayer networks, individuals can be actors on different layers
with different contacts in each layer. This is not necessarily the
case in interacting networks. Dickinson {\it et al.}
\cite{Dickison_12} studied numerically the SIR model in two networks
that interact through inter-layer connections given by a degree
distribution. There is a probability that these inter-layer
connections will allow infection to spread between nodes in different
layers. They found that, depending on the average degree of the
inter-layer connections, one layer can be in an epidemic-free phase
and the other in an epidemic phase. Yagan {\it et al.} \cite{Yag_13}
studied the SIR model in two multilayer networks in which all the
individuals act in both layers. In their model the transmissibility is
different in each network because one represents the virtual contact
network and the other the real contact network. They found that the
multilayer structure and the presence of the actors in both layers
make the propagation process more efficient and thus increase the
theoretical risk of infection above that found in isolated
networks. This can have catastrophic consequences for the healthy
population. Sanz {\it et al.} \cite{Sanz_14} studied the spreading
dynamics and the temporal evolution of two concurrent diseases that
interact with each other in a two-layer network system, for different
epidemic models. In particular, they found that for the SIR in the
final state this interaction can determinate the values of the
epidemic threshold of one of the diseases whose dynamic has been
modified by the presence of the other disease.  Buono {\it et al.}
\cite{Zuz_14} studied the SIR model, with $\beta$ and $t_r$
constant, in a system composed of two overlapping layers in which only
a fraction $q$ of individuals can act in both layers. In their model,
the two layers represent contact networks in which only the
overlapping nodes enable the propagation, and thus the
transmissibility $T$ is the same in both layers. They found that
decreasing the overlap decreases the transmissibility compared to when
there is a full overlap ($q=1$).

All of the above research assumes that individuals, independent of
their state, will continue acting in many layers. In a real-world
scenario, however, an infected individual may be isolated for a period
of time and thus may not be able to act in other layers, e.g., for a
period of time they may not be able to go to work or visit friends and
may have to stay at home or be hospitalized. Thus the propagation of
the disease is reduced. This scenario is more realistic than the one
in which an actor continues to participate in all layers irrespective
of their state \cite{Yag_13,Zuz_14}. As we will demonstrate, with our
approach the critical probability of infection is higher than the one
produced by the SIR model in a multilayer network.

\section*{RESULTS}

\subsection*{Model and Simulation Results}

We consider the case of a two-layer network, $A$ and $B$, of equal size
$N$, where one layer represents an individual's work environment and the
other their social environment. The degree distribution in each layer is
given by $P_i(k)$, with $i=A,B$ and $k_{\rm min} \leq k \leq k_{\rm
  max}$, where $k_{\rm min}$ and $k_{\rm max}$ are the minimum and the
maximum degree allowed a node.

At the initial stage of the Susceptible-Infected-Isolated-Recovered
model (SI$_{\rm I}$R) all individuals in both layers are susceptible
nodes. We randomly infect an individual in layer $A$. At the beginning
of the propagation process, each infected individual is isolated from
both layers with a probability $w$ for a period of time $t_w$. For
simplicity, in our epidemic model, we assume that every infected
individual is isolated from both layers with the same probability $w$
during a period of time $t_w$. The probability that an infected
individual is not isolated from both layers is thus $1-w$. At each
time step, a non-isolated infected individual spreads the disease with
a probability $\beta$ during a time interval $t_r$ after which he
recover. When an isolated individual $j$ after $t_w$ time steps
is no longer isolated he reverts to two possibles states. When $t_w <
t_r$, $j$ will be infected in both layers for only $t_r-t_w$ time
steps and the infection transmissibility of $j$ is reduced from
$1-(1-\beta)^{t_r}$ to $1-(1-\beta)^{t_r-t_w}$, but when $t_w \ge
t_r$, $j$ recovers and no longer spreads the disease. At the final
stage of the propagation all of the individuals are either susceptible
or recovered. The overall transmissibility $T^*\equiv
T^*_{\beta,t_r,t_w,w}$ is the probability that an infected individual
will transmit the disease to their neighbors. This probability takes
into account that the infected is either isolated or non-isolated in
both layers for a period of time and is given by
\begin{eqnarray}
 T^*&=&1-\left[(1-w)\;(1-\beta)^{t_r} +w\;(1-\beta)^{t_r-t_w}\right].
  \label{transmissibility}
\end{eqnarray}
Here the second and third term takes into account non-isolated and
isolated individuals and represents the probabilities that this
infected individual does not transmit the disease during $t_r$ and
$t_r-t_w$ time steps respectively.

\begin{figure}
  \centering
  \includegraphics[scale=0.135]{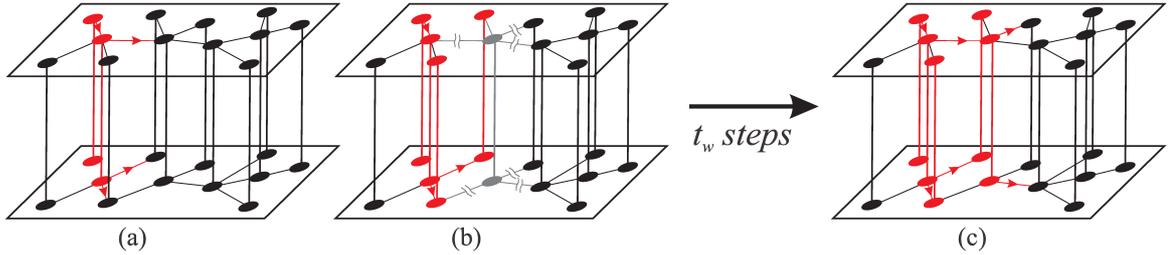}
  \caption{Schematic of a multilayer network consisting of two layers,
    each of size $N=12$. The black nodes represent the susceptible
    individuals and the red nodes the infected individuals. In this
    case, we consider $t_w < t_r$. (a) The red arrows indicate the
    direction of the branches of infection. All the branches spreads
    through $A$ and $B$ because the infected nodes are not isolated and
    thus interact in both layers. (b) The gray node, represents an
    individual who is isolated from both layers for a period of time
    $t_w$. (c) After $t_w$ time steps the gray node in (b) is no longer
    isolated, and can infect its neighbors in $A$ and $B$, if they were
    not reach by another branch of infection, during $t_r-t_w$ time
    steps (Color on line).}
  \label{ramasgraf}
\end{figure}

Mapping this process onto link percolation in two layers, we can write two
self-consistent coupled equations, $f_i$, $i=A,B$, for the probability
that in a randomly chosen edge traversed by the disease there will be a
node that facilitates an infinite branch of infection throughout the
two-layer network, i.e.,
\begin{eqnarray}
 f_A&=&[1-G_1^A \;(1-T^* f_A)\;G_0^B(1-T^* f_B)] \nonumber\\
 f_B&=&[1- G_1^B(1-T^* f_B) \;G_0^A(1-T^* f_A)],
  \label{ramas}
\end{eqnarray}
where $G_0^{A/B}$ and $G_1^{A/B}$ are the generating function defined
in the Introduction for layer $A$ and $B$. Here $G_1^{A/B}$ takes into
account the probability that a branch of infection reaches a node in
layer $A/B$ of connectivity $k$ across one of its links and cannot
expand through its remaining $k-1$ connection. Then $G_0^{A/B}$
represents the probability that the branch of infection propagates
from one layer into the other, reaches a node, but cannot expand
through all of its connections. Figure~(\ref{ramasgraf}) shows a
schematic of the contributions to Eqs.~(\ref{ramas}).

Using the nontrivial roots of Eq.~(\ref{ramas}) we
compute the order parameter of the phase transition, which is the
fraction of recovered nodes $R$, where $R$ is given by
\begin{equation}
 R=1-G_0^A \;(1-T^* f_A)\;G_0^B(1-T^* f_B).
  \label{recuperados}
\end{equation}
Note that in the final state of the process the fraction of recovered
nodes in layers $A$ and $B$ are equal because all nodes are present in
both layers. From Eqs.~(\ref{transmissibility}) and (\ref{ramas}) we
see that if we use the overall transmissibility $T^*$ as the control
parameter we lose information about $w$, the isolation parameter, and
$t_w$, the characteristic time of the isolation. In our model we thus
use $\beta \equiv \beta_{T^*}$ as the control parameter, where $\beta$
is obtained by inverting Eq.~(\ref{transmissibility}) with fixed
$t_r$. Notice that $\beta$ and $t_r$ are the intrinsic probability of
infection and recovery time of an epidemic obtained from epidemic
data. Thus making $t_r$ constant means that it is the average time of
the duration of the disease.

\begin{figure}
  \centering
  \includegraphics[scale=0.135]{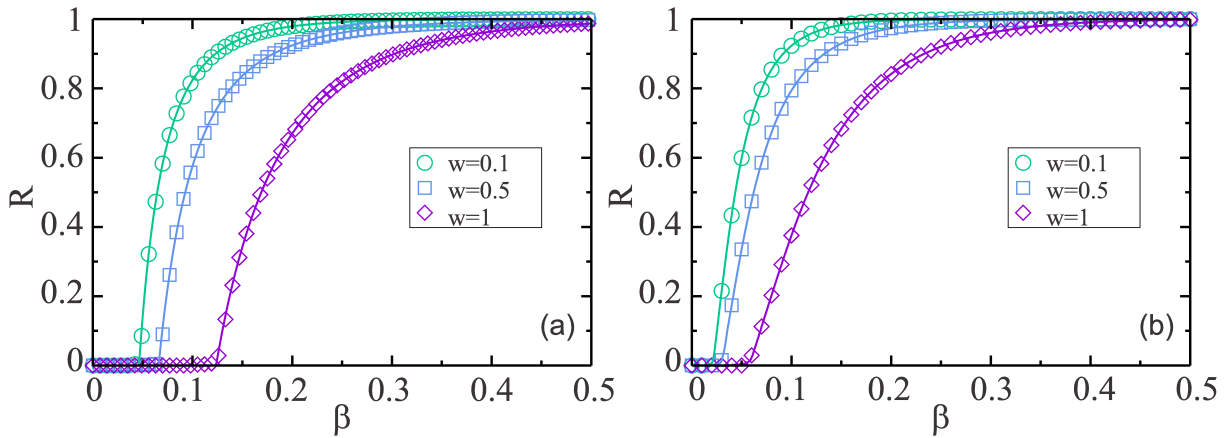}
  \caption{Simulations and theoretical results of the total fraction
    of recovered nodes $R$, in the final state of the process, as a
    function of $\beta$, with $t_r = 6$ and $t_w=4$, for different
    values of $w$. The full lines corresponds to the theoretical
    evaluation of Eq.~\ref{recuperados} and the symbols corresponds to
    the simulations results, for $w=0.1$ ($\bigcirc$) in green,
    $w=0.5$ ($\Box$) in blue and $w=1$ ($\Diamond$) in violet. The
    multilayer network is consisted by two layers, each of size
    $N=10^5$. For (a) two ER layers with $\langle k_A \rangle \simeq
    \langle k_B \rangle \simeq 2.31$, $k_{\rm min}=1$ and $k_{\rm
      max}=40$ and (b) two scale free networks with $\lambda_A=2.5$,
    $\lambda_B=3.5$ and exponential cutoff $c=20$ with $k_{\rm min}=2$
    and $k_{\rm max}=250$ (Color online).}
  \label{simulacion}
\end{figure}

Figure~\ref{simulacion} shows a plot of the order parameter $R$ as a
function of $\beta$ for different values of $w$, with $t_r=6$ and
$t_w=4$ obtained from Eq.~(\ref{recuperados}) and from the
simulations. For (a) we consider two Erd\H{o}s-R\'enyi (ER) networks
\cite{Erd_01}, which have a Poisson degree distribution and an average
degree $\langle k_A \rangle \simeq \langle k_B \rangle \simeq 2.31$,
and for (b) we consider two scale free networks with an exponential
cutoff $c=20$ \cite{New_03}, where $P_i(k_i) \sim k_i^{-\lambda_i}
e^{-k_i/c}$, with $\lambda_A=2.5$ and $\lambda_B=3.5$. We use this
type of SF network because it represents many structures found in
real-world systems \cite{Ama_01, Bata_00}.

In the simulations we construct two uncorrelated networks of equal
size using the Molloy-Reed algorithm \cite{Moll}, and we randomly
overlap one-to-one the nodes in network $A$ with the nodes of networks
$B$.  We assume that an epidemic occurs at each realization if the
number of recovered individuals is greater than $200$ for a system
size of $N=10^5$ \cite{Lag_02}. Realizations with fewer than $200$
recovered individuals are considered outbreaks and are disregarded.

\begin{figure}
  \centering
  \includegraphics[scale=0.135]{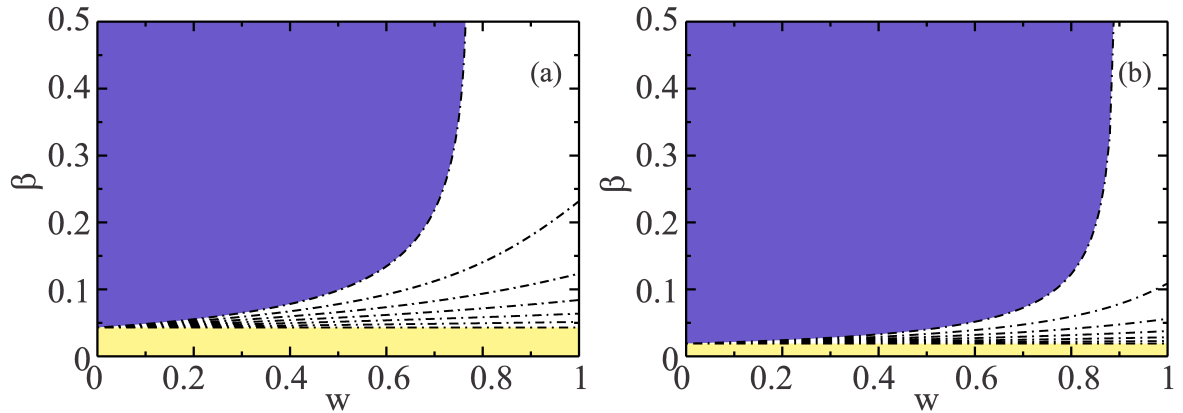}
  \caption{Phase diagram in the plane $\beta-w$. In both plots, we
  consider $t_r=6$ and $t_w=0,1,2,3,4,5,6$ from bottom to top for (a)
  two ER networks with $\langle k_A \rangle \simeq \langle k_B
  \rangle \simeq 2.31$ with $k_{\rm min}=1$ and $k_{\rm
    max}=40$. (b) two power law networks with $\lambda_A=2.5$ and
  $\lambda_B=3.5$ with $k_{\rm min}=2$ and $k_{\rm max}=250$ and
  exponential cutoff $c=20$. The region above each line corresponds to
  the Epidemic phase and the region below correspond to the
  Epidemic-free phase. In the limit of $w \to 0$ and for $t_w=0$ we
  recover the SIR in multiplex networks with (a) $\beta_c \approx
  0.043$ and (b) $\beta_c \approx 0.019$. For the case $t_r=t_w$,
  there is a threshold for $w$ with (a) $w_c=0.76$ and (b) $w_c=0.88$,
  above which there is only an Epidemic-free phase.}
  \label{diagrama}
\end{figure}

Figure~\ref{simulacion} shows an excellent agreement between the
theoretical equations (see Eq.~\ref{recuperados}) and the simulation
results. The plot shows that the critical threshold for an epidemic
$\beta_c$ increases with the isolation parameter $w$. Note that above
the threshold but near it $R$ decreases as the isolation $w$
increases, indicating that isolation for even a brief period of time
reduces the propagation of the disease. The critical threshold
$\beta_c$ is at the intersection of the two Eqs.~(\ref{ramas}) where
all branches of infection stop spreading, i.e., $f_A=f_B=0$. This is
equivalent to finding the solution of the system $det(J-I)=0$, where
$J$ is the Jacobian of the coupled equation with
$J_{i,k}|_{f_i=f_k=0}=\partial f_i / \partial f_k|_{f_i=f_k=0}$ and
$I$ is the identity, and
\begin{eqnarray}
T_c^{* \; 2} \left[(\kappa_A-1)(\kappa_B-1)-\langle k_A \rangle \langle k_B
  \rangle \right]-T_c^*[(\kappa_A-1)+(\kappa_B-1)]+1&=&0,
\label{resolvente}
\end{eqnarray}
where $\kappa_A$ and $\kappa_B$ are the branching factor of layers $A$
and $B$, and $\langle k_A \rangle$ and $\langle k_B \rangle$ are their
average degree. Using numerical evaluations of the roots of
Eq.~(\ref{resolvente}) we find the physical and stable solution for
the critical threshold $\beta_c$, which corresponds to the smaller
root of Eq.~(\ref{resolvente}) \cite{All_97}. Figure~\ref{diagrama}
shows a plot of the phase diagram in the plane $\beta-w$ for (a) two
ER multilayer networks \cite{Erd_01} with average degree $\langle k_A
\rangle \simeq \langle k_B \rangle \simeq 2.31$  and (b) two power law
networks with an exponential cutoff $c=20$ \cite{New_03}, with
$\lambda_A=2.5$ and $\lambda_B=3.5$. In both Fig.~\ref{resolvente} and
Fig.~\ref{diagrama} we use $t_r = 6$ and values $t_w=0$, $1$, $2$,
$3$, $4$, $5$, and $6$, from bottom to top.

The regions below the curves shown in Fig.~\ref{diagrama} correspond
to the epidemic-free phase. Note that for different values of $t_w$
those regions widen as $w$ increases. Note also that when $t_r = t_w$
there is a threshold $w_c$ above which, irrespective of the critical
epidemic threshold ($\beta_c)$, the disease never becomes an
epidemic. For $t_w=0$ and $w=0$ we recover the SIR process in a
two-layer network system that corresponds to $\beta_c \approx 0.043$
with $k_{\rm min}=1$ and $k_{\rm max}=40$ in Fig.~\ref{diagrama}(a)
and $\beta_c \approx 0.019$ with $k_{\rm min}=2$ and $k_{\rm max}=250$
in Fig.~\ref{diagrama}(b). Although in the limit $c \to \infty$,
$\beta_c \to 0$, most real-world networks are not that heterogeneous
and exhibit low values of $c$ \cite{New_05,Ama_01}.

As expected and confirmed by our model, the best way to stop the
propagation of a disease before it becomes an epidemic is to isolate
the infected individuals in both layers until they recover, which
corresponds to $t_w=t_r$ and $w > 0$. Because this is strongly
dependent upon the resources of the location from which the disease
begins to spread and on each infected patient's knowledge of the
consequences of being in contact with healthy individuals, the
isolation procedure can be difficult to implement.

We also study a case in which there is isolation in only one layer
(for a detailed description see Supplementary Information). We find
that there is no critical value $w_c$ above which the phase is
epidemic-free, i.e., above $\beta_c$ and for all values of $w$ the
disease always becomes an epidemic.

\begin{figure}
  \centering
  \includegraphics[scale=0.135]{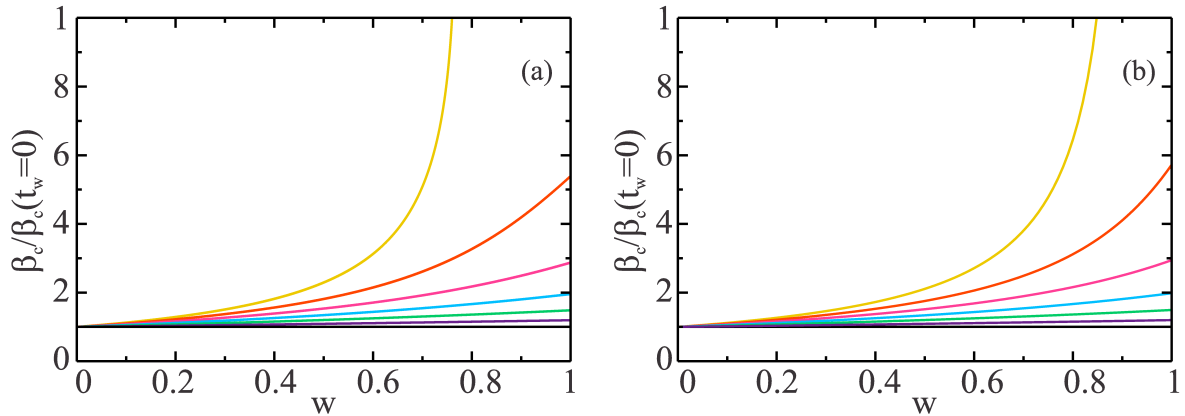}
  \caption{Ratio of $\beta_c(t_w)$ to $\beta_c(0)$ as a function of
    $w$. For $t_w=1,2,3,4,5,6$ from bottom to top for (a) two ER
    networks with $\langle k_A \rangle \simeq \langle k_B \rangle \simeq 2.31$
    with $k_{\rm min}=1$ and $k_{\rm max}=40$ and (b) two power law
    networks with $\lambda_A=2.5$ and $\lambda_B=3.5$ with $k_{\rm
      min}=2$ and $k_{\rm max}=250$, with exponential cutoff
    $c=20$. In both Figures, the limit $w \to 0$ correspond to a SIR
    process, and as $w$ increases the underestimation increases.}
  \label{sobreestimacion}
\end{figure}

The phase diagram indicates that when the SIR model is applied to
multilayer networks, which corresponds to the case $t_w=0$, it
underestimates the critical threshold $\beta_c$ of an epidemic. This
underestimation can strongly affect the spreading dynamics.
Figure~\ref{sobreestimacion}(a) plots the ratio
$\beta_c/\beta_c(t_w=0)$ as a function of $w$ for different values of
$t_w$, with $t_w > 0$ for two ER
networks. Figure~\ref{sobreestimacion}(b) shows how much more the critical
threshold is underestimated in the SIR model of two-layer SF
networks than in the SI$_{\rm I}$R model.

In the limit $t_w = 0$ and $w \to 0$ we revert to the SIR model in
multilayer networks~\cite{Zuz_14}. As $w$ increases and when $t_w \neq
0$ there is always an underestimation of the critical threshold. Note
that when $t_w=t_r$ the plot shows that when the percentage of
infected individuals who are hospitalized or isolated in their homes
is approximately $40\%$, for two ER, and $50\%$, for two SF, the
SI$_{\rm I}$R model indicates double the actual critical threshold of
infection than that indicated in the SIR model. The declaration of an
epidemic by a government health service is a non-trivial decision, and
can cause panic and negatively effect the economy of the region. Thus
any epidemic model of airborne diseases that spread in multilayer
networks, if the projected scenario is to be realistic and in
agreement with the available real data, must take into account that
some infected individuals will be isolated for a period of time. Note
that isolation can represent behavioral change but, unlike previous
models in which the behavioral changes are solely the result of
decisions made by susceptible individuals \cite{Funk,Perra}, our model
allows behavioral changes brought about by placing the infected
individuals in quarantine or by hospitalizing them
\cite{Leg_06,Gomes_14,Cai_14,Val_15}, two practices that were
instituted during the recent Ebola outbreak in West Africa.  Also note
that this isolation can delay the onset of the peak of the epidemic
and thus allow health authorities more time to make
interventions. This is an important topic for future investigation.

\section*{DISCUSSION}  

In summary, we study a SI$_{\rm I}$R epidemic model in a two-layer
network in which infected individuals are isolated from both layers
with probability $w$ during a period of time $t_w$. Using the
framework of link percolation based on a generating function, we
compute the total fraction of recovered nodes in the steady state as a
function of the probability of infection $\beta$ and find a perfect
agreement between the theoretical and the simulation results. We
derive an expression for the intrinsic epidemic threshold and we find
that $\beta_c$ increases as $w$ and $t_w$ increase. For $t_w=t_r$ we
find a critical threshold $w_c$ above which any disease never becomes
an epidemic and which cannot be found when isolating only in one
layer. From our results we also note that as the isolation parameter
and the period of isolation increases the underestimation
increases. Our model enables us to conclude that the SIR model of
multilayer networks without isolation underestimates the critical
infection threshold. Thus the isolation of the infected individuals,
in both layers, for a period of time should be included in future
epidemic models in which individuals can recover.

\bigskip

\noindent{\it Acknowledgments}

\noindent We thank the NSF (grants CMMI 1125290 and CHE-1213217) and
the Keck Foundation for financial support. LGAZ and LAB wish to thank
to UNMdP and FONCyT (Pict 0429/2013) for financial support.

\bigskip

\noindent{\it Additional information} 

\noindent The authors declare no competing financial interests.
Supplementary information is available in the online version of the
paper.  Reprints and permissions information is available online at
www.nature.com/reprints. Correspondence and requests for materials
should be addressed to LGAZ.

\bigskip

\noindent{\it Author Contribution Statement}

\noindent L.G.A.Z. and L.A.B. wrote the main manuscript text and L.G.A.Z.
prepared figures 1-4. All authors performed the research and reviewed
the manuscript.

\bigskip

\newpage

\title{ Epidemic Model with Isolation in Multilayer Networks}

\author{L. G. Alvarez Zuzek} \email{lgalvere@mdp.edu.ar}
\affiliation{Departamento de F\'{i}sica, Facultad de Ciencias Exactas
  y Naturales, Universidad Nacional de Mar del Plata, and Instituto de
  Investigaciones F\'{\i}sicas de Mar del Plata (IFIMAR-CONICET),
  De\'an Funes 3350, 7600 Mar del Plata, Argentina}
\author{H. E. Stanley} \affiliation{Center for Polymer Studies, Boston
  University, Boston, Massachusetts 02215, USA.}
\author{L. A. Braunstein} \affiliation{Departamento de F\'{i}sica,
  Facultad de Ciencias Exactas y Naturales, Universidad Nacional de
  Mar del Plata, and Instituto de Investigaciones F\'{\i}sicas de Mar
  del Plata (IFIMAR-CONICET), De\'an Funes 3350, 7600 Mar del Plata,
  Argentina}\affiliation{Center for Polymer Studies, Boston
  University, Boston, Massachusetts 02215, USA.}

\maketitle

\section*{{\large Supplementary Information}}

We here compare the critical values of our model in which the
isolation occurs in both layers, with the critical values obtained by
a model in which the isolation occurs in only one layer.  This
situation could reflect a real-world scenario in which infected
individuals do not go to their jobs, in one layer, but still have
contact with people, in the other layer. At the initial stage all the
{\bf $N$} individuals in both layers are susceptible and we randomly
infect a node in layer $A$. With a probability $w$ this node is
isolated in layer $A$, but it is not isolated in layer $B$ and can
spread the disease there. During the disease spreading the isolated
infected nodes in layer $A$ spread the disease for a shorter period of
time than the non-isolated nodes. Thus the transmissibility of
isolated individuals in layer $A$ is $1-(1-\beta)^{t_r-t_w}$, and the
transmissibility of non-isolated individuals in $A$ and all infected
individuals in $B$ is $1-(1-\beta)^{t_r}$. At the final stage of the
propagation all individuals are either susceptible or recovered, and
the transmissibilities $T^A$ and $T^B$ in layer $A$ and $B$
respectively are
\begin{eqnarray}
  T^A&=&1-\left[(1-w)\;(1-\beta)^{t_r} +w\;(1-\beta)^{t_r-t_w}\right],
  \nonumber\\ 
  T^B&=&1-\left[(1-\beta)^{t_r}\right],
  \label{transmissibility2}
\end{eqnarray}
where $(1-w)\;(1-\beta)^{t_r}$ is the probability that a non-isolated
infected individual will not transmit the disease for a period of time
$t_r$ in layer $A$, $w\;(1-\beta)^{t_r-t_w}$ is the probability that
an infected isolated individual in layer $A$ will not transmit the
disease during $t_r-t_w$ time steps, and $(1-\beta)^{t_r}$ is the
probability that an infected individual will not transmit the disease
until they recover after $t_r$ time steps since they were infected.

Using the theoretical arguments presented in Model and Simulation
Results, we write two self-consistent coupled equations for the
probability that the branches of infection will expand an infinite
cluster of recovered individuals at the final stage of the
propagation,
\begin{eqnarray}
 f_A&=&[1-G_1^A \;(1-T^A f_A)\;G_0^B(1-T^B f_B)] \nonumber\\
 f_B&=&[1- G_1^B(1-T^B f_B) \;G_0^A(1-T^A f_A)].
  \label{ramas2}
\end{eqnarray}
The critical threshold $\beta_c$ is at the intersection of the two
Eqs.~(\ref{ramas2}) where all branches of infection do not spread, i.e.,
$f_A=f_B=0$. This is equivalent to finding the solution of the system
$det(J-I)=0$, where $J$ is the Jacobian of the coupled equation with
$J_{i,k}|_{f_i=f_k=0}=\partial f_i / \partial f_k|_{f_i=f_k=0}$ and $I$
is the identity (See Model and Simulation Results in the main text),
from where
\begin{eqnarray}
(T_c^{A} T_c^{B})^2 \left[(\kappa_A-1)(\kappa_B-1)-\langle k_A \rangle \langle k_B
  \rangle \right]-T_c^A(\kappa_A-1)-T_c^B(\kappa_B-1)+1&=&0.
\label{resolvente2}
\end{eqnarray}
Here $\kappa_A$ and $\kappa_B$ are the branching factors of layers $A$
and $B$ respectively, and $\langle k_A \rangle$ and $\langle k_B
\rangle$ are their average degree. The physical and stable solution
for the critical threshold $\beta_c$ corresponds to the smaller root
of Eq.~(\ref{resolvente2}).

\begin{figure}
  \centering
  \includegraphics[width=0.95\textwidth]{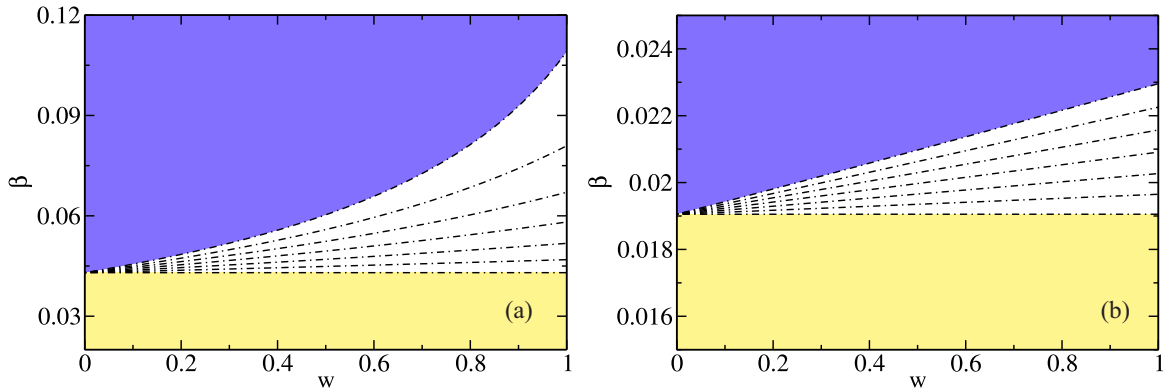}
\caption{Phase diagram in the plane $\beta-w$. In both plots, we
  consider $t_r=6$ and $t_w=0,1,2,3,4,5,6$ from bottom to top for (a)
  two ER networks with $\langle k_A \rangle \simeq \langle k_B \rangle
  \simeq 2.31$ with $k_{\rm min}=1$ and $k_{\rm max}=40$. (b) two
  power law networks with $\lambda_A=2.5$ and $\lambda_B=3.5$ with
  $k_{\rm min}=2$ and $k_{\rm max}=250$ and exponential cutoff
  $c=20$. The region above each line corresponds to the Epidemic phase
  and the region below correspond to the Epidemic-free phase. In the
  limit of $w \to 0$ and for $t_w=0$ we recover the SIR in multiplex
  networks with (a) $\beta_c \approx 0.043$ and (b) $\beta_c \approx
  0.019$.}
  \label{diagrama2}
\end{figure}

Figure \ref{diagrama2} shows a plot of the phase diagram in the plane
$\beta-w$ for (a) two ER multilayer networks \cite{Erd_01} with
average degree $\langle k_A \rangle \simeq \langle k_B \rangle \simeq 2.31$ and
(b) two power-law networks with an exponential cutoff $c=20$
\cite{New_03}, with $\lambda_A=2.5$ and $\lambda_B=3.5$. In
Fig.~\ref{diagrama2} we use $t_r = 6$ and values $t_w=0$, 1, 2, 3, 4,
5, and 6, from bottom to top. 

Note that the regions below the curves in Fig.~\ref{diagrama2}
correspond to the epidemic-free phase. For different values of
$t_w$ these regions than widen as $w$ increases and reach their maximum
size for $t_r$ equal to $t_w$. 

\begin{figure}
  \centering
  \includegraphics[width=0.85\textwidth]{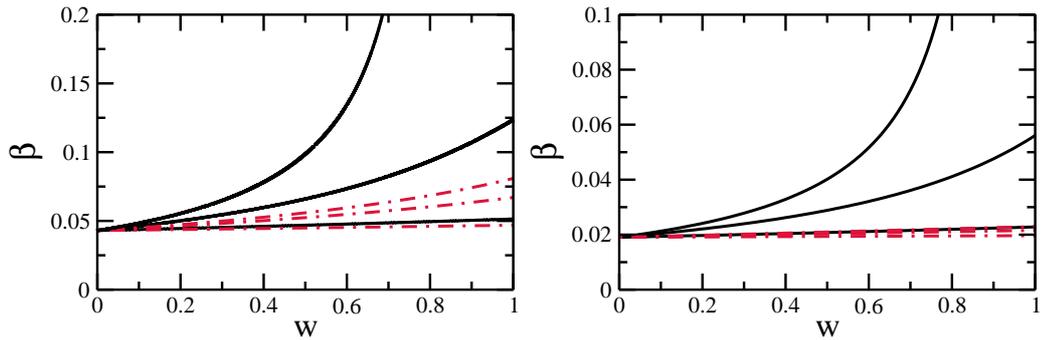}
  \caption{Phase diagram in the plane $\beta-w$. We consider $t_r=6$
    and $t_w=1,4,6$ from bottom to top. The black lines correspond to
    isolation in both layers and the red lines correspond to the
    scenario presented in this section. For (a) two ER networks with
    $\langle k_A \rangle \simeq \langle k_B \rangle \simeq 2.31$ with
    $k_{\rm min}=1$ and $k_{\rm max}=40$. (b) two power law networks
    with $\lambda_A=2.5$ and $\lambda_B=3.5$ with $k_{\rm min}=2$ and
    $k_{\rm max}=250$ and exponential cutoff $c=20$.}
  \label{comparacion}
\end{figure}

In order to compare the two scenarios in which isolation takes place in
both layers or in one layer, in Fig.~ \ref{comparacion} we plot the
phase diagram in the plane $\beta-w$ in both situations for $t_w=1$, 4,
and 6, with $t_r=6$.

As expected in the scenario where isolation takes place in one network,
the plot shows that as $t_w$ and $w$ increases the critical values of
$\beta$ increase but much less than in the scenario where isolation is
considered in both layers. For example, for two ER networks when $w=0.5$
and $t_w=1$ both scenarios have approximately the same value of $\beta$,
but when $w=0.5$ and $t_w=6$ the value of $\beta$ increases twofold
comparing one scenario with the other. When $w>0.8$, in the scenario
presented in this section we find that above some values of $\beta$
there is an epidemic-phase. In contrast, in the earlier scenario with
isolation in both layers there is no value of $\beta$ above which there
is an epidemic-phase, i.e., the spreading disease never becomes an
epidemic. When there are two SF layers there is an even more extreme
behavior: $\beta$ varies slightly with $w$ in the scenario with
isolation in one layer, but when there is isolation in both layers we
find a threshold above which there is no epidemic-phase.


\bigskip

\begin{thebibliography}{99}

\bibitem{Bara_rew} Albert R. \& Barab{\'a}si A. L.  Statistical
  mechanics of complex networks. {\it Rev. Mod. Phys.} {\bf 74}, 47
  (2002).

\bibitem{Lar_09} La Rocca C. E., Braunstein L. A. \& Macri
  P. A. Conservative model for synchronization problems in complex
  netwrks. {\it Phys. Rev. E.} {\bf 80}, 26111 (2009).

\bibitem{Ana_01} Pastore y Piontti A., Macri P. A. \& Braunstein
  L. A. Discrete surface growth process as a synchronization mechanism
  for scale free complex networks. {\it Phys. Rev. E.} {\bf 76},
  46117 (2007).

\bibitem{Pan_04} Gallos L. K. \& Argyrakis P. Absence of kinetic
  effects in reaction-diffusion processes in scale-free networks.
  {\it Phys. Rev. Lett.} {\bf 92},b138301 (2004).

\bibitem{Dun_01} Callaway D. S., Newman M. E. J., Strogatz S. H.
 \& Watts D. J. Network Robustness and Fragility: Percolation on
  Random Graphs. {\it Phys. Rev. Lett.} {\bf 85}, 5468 (2000).

\bibitem{Coh_03} Cohen R., Havlin S. \& Ben-Avraham D. Efficient
  Immunization Strategies for Computer Networks and Populations. {\it
    Phys. Rev. Lett.} {\bf 91}, 247901 (2003).

\bibitem{New_03} Newman M. E. J., Strogatz S. H. \& Watts
  D. J. Random graphs with arbitrary degree distributions and their
  applications. {\it Phys. Rev. E} {\bf 64}, 26118 (2001).

\bibitem{Val_11} Valdez L. D., Buono C., Macri P. A. \& Braunstein
  L. A. Effect of degree correlations above the first shell on the
  percolation transition. {\it Europhysics lett.} {\bf 96}, 38001
  (2011).

\bibitem{New_05} Newman M. E. J. Spread of epidemic disease on
  networks. {\it Phys. Rev. E.} {\bf 66}, 16128 (2002).

\bibitem{past_01} Pastor-Satorras R. \& Vespignani A. Epidemic
  Spreading in Scale-Free Networks. {\it Phys. Rev. Lett.} {\bf 86},
  3200 (2001).

\bibitem{Buo_13} Buono C., Vazquez F., Macri P. A. \& Braunstein
  L. A. Slow epidemic extinction in populations with heterogeneous
  infection rates. {\it Phys. Rev. E.} {\bf 88}, 22813 (2013).

\bibitem{past_02} Pastor-Satorras R. \& Vespignani A. Epidemic
  dynamics and endemic states in complex networks. {\it Phys. Rev. E.}
  {\bf 63}, 66117 (2001).

\bibitem{Gra13} Granell C., G{\'o}mez S. \& Arenas A. Dynamical
  Interplay between Awareness and Epidemic Spreading in Multiplex
  Networks. {\it Phys. Rev. Lett.} {\bf 111}, 128701 (2013).

\bibitem{Cozzo_13} Cozzo E., Ba{\~n}os R. A., Meloni S. \& Moreno
  Y. Contact-based Social Contagion in Multiplex Networks. {\it
    Phys. Rev. E.} {\bf 88}, 50801(R) (2013).

\bibitem{Mar_11} Marceau V., No{\"e}l P. A., H{\'e}bert-Dufresne
  L., Allard A. \& Dub{\'e} L. J. Modeling the dynamical interaction
  between epidemics on overlay networks. {\it Phys. Rev. E.} {\bf
    84}, 26105 (2011).

\bibitem{Sanz_14} Sanz J., Xia C., Meloni S. \& Moreno Y. Dynamics of
  Interacting Diseases. {\it Phys. Rev. X.} {\bf 4}, 41005 (2014).

\bibitem{Sahneh_14} Sahneh F. D. \& Scoglio C. Competitive Epidemic
  Spreading Over Arbitrary Multilayer Networks. {\it Phys. Rev. E.}
  {\bf 89}, 62817 (2014).

\bibitem{Ander_91}Anderson, R. M., \& May, R. M. Infectious
  diseases of humans. {\it Oxford university press} {\bf 1} (1991).

\bibitem{Bai_75} Bailey N. T. The mathematical theory of infectious
  diseases and its applications. Griffin, London (1975).

\bibitem{Col_07} Colizza V., Barrat A., Barth{\'e}lemy M. \&
  Vespignani A. Predictability and epidemic pathways in global
  outbreaks of infectious diseases: the SARS case study. {\it BMC
    Medicine} {\bf 5}, 34 (2007).

\bibitem{Coh_hand} Cohen R., Havlin S. \& Ben-Avraham D. {\it
  Handbook of graphs and networks} (Wiley-VCH, Berlin, 2002),
  chap. Structural properties of scale free networks.

\bibitem{Bra_07} Braunstein L. A., {\it et al.}.  Optimal path and
  minimal spanning trees in random weighted networks. {\it Bifurcation
    and Chaos} {\bf 17}, 2215 (2007).

\bibitem{Bul_01} Buldyrev S. V., Parshani R., Paul G. \& Stanley
  H. E., Havlin S. Catastrophic cascade of failures in
  interdependent networks. {\it Nature} {\bf 464}, 1025 (2010).

\bibitem{jia_02} Gao J., Buldyrev S. V., Havlin S. \& Stanley
  H. E. Robustness of a Network of Networks. {\it Phys. Rev. Lett.}
  {\bf 107}, 195701 (2011).

\bibitem{Gao_12} Gao J., Buldyrev S. V., Stanley H. E \& Havlin
  S. Networks Formed from Interdependent Networks. {\it Nature
    Physics} {\bf 8}, 40 (2012).

\bibitem{Gao_01} Dong G., {\it et al.}. Robustness of network of
  networks under targeted attack. {\it Phys. Rev. E.} {\bf 87}, 52804
  (2013).

\bibitem{Val13} Valdez L. D., Macri P. A. \& Braunstein L. A. Triple
  point in correlated interdependent networks. {\it Phys. Rev. E.}
  {\bf 88}, 50803(R) (2013).
  
\bibitem{Bax_01} Baxter G. J., Dorogovtsev S. N., Goltsev A. V.  \&
  Mendes J. F. F. Avalanche Collapse of Interdependent Networks. {\it
    Phys. Rev. Lett.} {\bf 109}, 248701 (2012).

\bibitem{Bru_01} Brummitt C. D., D'Souza R. M. \& Leicht
  E. A. Suppressing cascades of load in interdependent networks. {\it
    Proceedings of the National Academy of Sciences} {\bf 109}, 680
  (2012).

\bibitem{Brummitt_12} Brummitt C. D., Lee K.-M. \& Goh
  K.-I. Multiplexity-facilitated cascades in networks. {\it
    Phys. Rev. E.} {\bf 85}, 45102(R) (2012).

\bibitem{Lee_12} Lee K.-M.,  Kim Jung Yeol, Cho W. K.,
   Goh K.-I. \& Kim  I.-M. Correlated multiplexity and connectivity of
  multiplex random networks. {\it New Journal of Physics} {\bf 14},
  33027 (2012).

\bibitem{Gomez_13} G{\'o}mez S., {\it et al.}. Diffusion Dynamics on
  Multiplex Networks. {\it Phys. Rev. Lett.} {\bf 110}, 28701 (2013).

\bibitem{Kim_13} Kim J. Y. \& Goh K.-I. Coevolution and Correlated
  Multiplexity in Multiplex Networks. {\it Phys. Rev. Lett.} {\bf
    111}, 58702 (2013).

\bibitem{Cozzo_12} Cozzo E., Arenas A. \& Moreno Y. Stability of
  Boolean multilevel networks. {\it Phys. Rev. E.} {\bf 86}, 36115
  (2012).

\bibitem{Car_02} Cardillo A., {\it et al.}. Emergence of Network
  Features from Multiplexity. {\it Scientific Reports} {\bf 3}, 1344
  (2013).

\bibitem{Kal_13} Kaluza P., K{\"o}lzsch A., Gastner M. T. \&
  Blasius B. The complex network of global cargo ship movements. {\it
    Journal of the Royal Society: Interface} {\bf 7}, 1093 (2010).

\bibitem{Dickison_12} Dickison M., Havlin S. \& Stanley
  H. E. Epidemics on interconnected networks. {\it Phys. Rev. E.} {\bf
    85}, 66109 (2012).

\bibitem{Yag_13} Yagan O., Qian D., Zhang J. \& Cochran
  D. Conjoining Speeds up Information Diffusion in Overlaying
  Social-Physical Networks. {\it IEEE JSAC Special Issue on Network
    Science} {\bf 31}, 1038 (2013).

\bibitem{Zuz_14} Buono C., Alvarez Zuzek L. G., Macri P. A \&
  Braunstein L. A. Epidemics in partially overlapped multiplex
  networks. {\it PLOS ONE} {\bf 9}, e92200 (2014).

\bibitem{Erd_01} Erd\H{o}s P. \& R{\'e}nyi A. On Random
  Graphs. I. {\it Publications Mathematicae} {\bf 6}, 290 (1959).

\bibitem{Ama_01} Amaral L. A. N., Scala A., Barth{\'e}lemy M. \&
  Stanley H. E. Classes of Small-World Networks. {\it
    Proc. Natl. Acad. Sci. USA} {\bf 97}, 11149 (2000).

\bibitem{Bata_00} Batagelj V., \& Mrvar A. Some analyses of Erdos
  collaboration graph. {\it Social networks} {\bf 22}, 173 (2000).

\bibitem{Moll} Molloy M \& Reed B. A critical point for random graphs
  with a given degree sequence. {\it Random Structures and Algorithms}
  {\bf 6}, 161 (1995).

\bibitem{Lag_02} Lagorio C., Migueles M. V., Braunstein L. A.
 , L{\'o}pez E. \& Macri P. A. Effects of epidemic threshold
  definition on disease spread statistics. {\it Physica A.} {\bf 388},
  755 (2009).

\bibitem{All_97} Alligood K. T., Sauer T. D. \& Yorke J. A. CHAOS: An
  Introduction to Dynamical Systems. {\it Springer} (1997).


\bibitem{Funk} Funk S., Salathe M \& Jansen V. A. A. .Modelling the
  influence of human behaviour on the spread of infectious diseases: a
  review {\it Journal of The Royal Society Interface} {\bf 7} (50),
  1247 (2010).

\bibitem{Perra} Perra N, Balcan D, Goncalves B, Vespignani A. Towards
  a Characterization of Behavior-Disease Models. {\it PLoS ONE} 6(8):
  e23084. doi:10.1371/ journal.pone.0023084 (2011).

\bibitem{Leg_06} Legrand J., {\it et al.}  Understanding the dynamics of Ebola
  epidemics. {\it PLOS Current Outbreaks} {\bf 135}, 610 (2006).

\bibitem{Gomes_14} Gomes M. F. C., {\it et al.}. Assessing the
  International Spreading Risk Associated with the 2014 West African
  Ebola Outbreak. {\it PLOS Current Outbreaks}
  doi:10.1371/currents.outbreaks.cd818f63d40e24aef769dda7df9e0da5
  (2014).

\bibitem{Cai_14} Rivers C. M. {\it et al.} Modeling the Impact of
  Interventions on an Epidemic of Ebola in Sierra Leone and
  Liberia. {\it PLOS Current Outbreaks}
  doi:10.1371/currents.outbreaks.4d41fe5d6c05e9df30ddce33c66d084c
  (2014).

\bibitem{Val_15} Valdez, L. D., Rêgo, H. H. A., Stanley, H. E., \&
  Braunstein, L. A. Predicting the extinction of Ebola
  spreading in Liberia due to mitigation strategies. arXiv preprint
  arXiv:1502.01326 (2015).


\end{thebibliography}

\bigskip

\begin{thebibliography}{99}



\bibitem{Erd_01} Erd\H{o}s P. \& R{\'e}nyi A. On Random
  Graphs. I. {\it Publications Mathematicae} {\bf 6}, 290 (1959).

\bibitem{New_03} Newman M. E. J., Strogatz S. H. \& Watts
  D. J. Random graphs with arbitrary degree distributions and their
  applications.  {\it Phys. Rev. E} {\bf 64}, 026118 (2001).




\end{thebibliography}
\end{document}